\documentclass[twocolumn,aps,prc]{revtex4}

\usepackage[utf8]{inputenc}
\usepackage{amsfonts}
\usepackage{amsmath}
\usepackage{amssymb}
\usepackage{epsfig}
\usepackage{bm}

\usepackage{fixme}
\usepackage{color}
\usepackage{cancel}
\begin{document}

\title{Two-body continuum states in non-integer geometry}

\author{E.R. Christensen$^{2}$, E. Garrido$^{1}$, A.S. Jensen$^{2}$}

\affiliation{$^{1}$Instituto de Estructura de la Materia, IEM-CSIC,
Serrano 123, E-28006 Madrid, Spain}

\affiliation{$^{2}$Department of Physics and Astronomy, Aarhus University, DK-8000 Aarhus C, Denmark}

\date{\today}

\begin{abstract}
Wave functions, phase shifts and corresponding elastic cross sections
are investigated for two short-range interacting particles in a
deformed external oscillator field.  For this we use the equivalent
$d$-method employing a non-integer dimension $d$.  Using a square-well
potential, we derive analytic expressions for scattering lengths and
phase shifts. In particular, we consider the dimension, $d_E$, for
infinite scattering length, where the Efimov effect occurs by addition
of a third particle.  We give explicitly the equivalent continuum wave
functions in $d$ and ordinary three dimensional (3D) space, and show that the phase shifts
are the same in both methods. Consequently the $d$-method can be used
to obtain low-energy two-body elastic cross sections in an external
field.
\end{abstract}


%21.45.-v : Few-body systems

%21.60.n : Nuclear structure models and methods

\maketitle

\section{Introduction}                            
\label{sectI}

The experimental possibilities with cold atomic gases allow both huge
two-body interaction variation as well as an overall confining
deformed external field \cite{koh06,blo08,chi10,den16}.  Few-body
physics quickly becomes complicated or even impossible as the number
of particles increase, and other techniques have to be employed, see
for example \cite{hov18}.  One purpose of studying these cold systems
is the experimental availability, where laboratory simulation and
manipulation are possible.  These systems may themselves be of
practical use, but also able to teach us how to control and create
similar properties in systems, which so far are outside of experimental
reach.  Comparing universal behavior of systems from different
subfields of physics, perhaps also chemistry and mathematics, is then
a tool to exchange knowledge between science subfields
\cite{efi70,sim76,lan77,nie01,jen04,fre12,zin13,nai17,gar18}.

The advantage of few-body physics is that all relevant
degrees-of-freedom can be accurately treated.  Increasing the number
of particles from two to three is known to increase substantially the
complications, like the appearance of new possible structures, 
but also the interest.  The
complications are accentuated by use of overall external fields, where
the description requires more than relative degrees-of-freedom.  It is
then worth cashing in on a reduction of degrees-of-freedom, which is
very useful, by means of the use of an effective dimension that changes continuously
\cite{nie01,lev14,yam15,san18,ros18}.

We shall in this report concentrate on the $d$-method, introduced in
Ref.~\cite{nie01}, and subsequently developed and applied for two and
three particles \cite{chr18,gar19a,gar19b,gar20,gar21}.  The
theoretical formulation assumes non-integer dimensions, $d$, only
relative degrees of freedom, but with a $d$-dependent angular momentum barrier,
and spherical calculations without angular dependence corresponding to zero 
angular momentum for $d=3$.
The equivalence to the description with an external deformed oscillator
potential is available, such that the $d$-method results can be translated
to ordinary three-dimensional calculations, and in this way open for
experimental comparison \cite{gar21}.

Interesting features revealed and highlighted by this method can be
understood from the continuous change of dimension from $3$ to $2$.
The properties of quantum solutions for simple two-body systems differ
enormously between these two dimensions.  For $d=3$, even for
relative $s$-waves, a finite attraction is necessary to bind, whereas for $d=2$ 
and relative $s$-waves, an infinitely small
attraction produces a bound state \cite{bru79,lim80,nie97,vol13,nai17}.
For three particles the differences are much larger, as emphasized by
the existing Efimov effect for $d=3$, defined by no bound two-body
subsystems, but infinitely many bound three-body systems.  In
contrast, only a finite number of bound states exist for $d=2$.
Thus varying $d$ between $3$ and $2$ necessarily implies that some
states must change behavior from bound to continuum states or vice
versa.  One especially spectacular phenomenon is that the Efimov
effect can be induced between dimensions $3$ and $2$, while absent in
both ends of the interval \cite{gar21}.

The $d$-variation amounts to varying an external field, which is a
familiar procedure.  This is exemplified from the Feshbach tuning of a
magnetic field allowing huge changes of the effective two-body
interaction \cite{vog97,cou98}.  This is the same principle as for
variation of an external electric field as first suggested in
Ref.~\cite{nie99}.  The present use of the $d$-method can then be
viewed as a theoretical application of an external field without the
necessary additional degrees-of-freedom.  However, these fields are
unavoidable when comparing with practical experiments.

Continuum or scattering calculations are usually more difficult than
bound state investigations \cite{sha18}.  It is therefore tempting to
extend the $d$-method to the continuum, and first to learn from the
simplest system, that is, a two-body system. The task is then to
perform calculations with the $d$-method, and subsequently translate
or interpret the results to three dimensions with an external field.
We shall study two particles
interacting through simple short-range potentials.  The main
properties can then be revealed by square-well potentials, from which
we can derive analytic solutions with characteristic universal
behavior. These square-well results may be compared to numerically
found solutions for more realistic Gaussian potentials.

The purpose of the present report is therefore to study continuum
properties of two-body systems with the $d$-method, and interpret as if
obtained in an external field with the possible experimental
comparison. Special attention will be paid to the dimension, $d=d_E$, for
which the two-body binding energy is equal to zero. This dimension is
particularly interesting, since for $d=d_E$ the Efimov conditions are fulfilled 
if a third particle is added.
 
After the introduction in Section~\ref{sectI}, we derive in Section~\ref{sectII}
several analytical expressions of interest assuming a square-well interaction.
In the different subsections we provide the energy of bound states as a function 
of $d$ and of the strength of the potential, the critical dimension $d_E$
as a function of the strength of the potential, and the phase shifts also as a function of 
$d$ and of the strength of the potential. Finally, we also discuss in this 
section how these phase
shifts have to be treated in order to construct the cross sections.
The results and illustrations are provided in Section~\ref{sectIII},
where together with the direct applications of the expressions derived
in Section~\ref{sectII}, we focus on the universality of the results, 
investigating how the values of $d_E$ and the phase shifts can be shown
to present a universal behavior, independent of the details of the potential.
We finish wit Section \ref{sectV}, which contains the summary and
conclusions. Pertinent properties of the Bessel functions are
collected in an appendix.

\section{Two-body scattering}
\label{sectII}

We consider the two-body problem in $d$ dimensions,
where $2 \leq d \leq 3$, and where the two-body interaction is assumed 
to be of short range. Using a schematic short-range
square-well potential, we shall derive expressions for bound- and continuum-state
energies, threshold conditions, phase shifts, scattering lengths, and
cross sections, all as a function of $d$.  In the end, these
results can be interpreted as obtained by an external deformed field
and full three-dimensional space.

The derived analytic properties, properly expressed, exhibit the highly 
desired universal behavior. This universality means potential independence 
at distances larger than the radius of the short-range interaction. In other
words, gross properties are sufficient to describe the results for relatively small
energies. Numerical calculations using a Gaussian potential will be used
as an illustration.

\subsection{Theoretical formulation}

For two-body systems and dimension $d$, the reduced, $R(r)$, and the total, $\psi(r)$, radial 
relative wave functions are related by $R(r) = r^{(d-1)/2} \psi(r)$, where $r$ is
the relative distance coordinate. The reduced wave function, $R(r)$, is obtained
as a solution, with the proper boundary conditions, of the $d$-dimensional radial 
Schr\"{o}dinger equation for the relative motion \cite{nie01}: 
\begin{eqnarray} \label{e400}
  \left[ - \frac{\hbar^2}{2\mu} \bigg(\frac{\partial^2}{\partial r^2}
  - \frac{\ell^*(\ell^*+1)}{r^2} \bigg) +  V(r)-E \right] R(r)=0  \; ,
\end{eqnarray}
where $V(r)$ is the spherical potential and $\mu$ the reduced mass. 
For two-body systems the effective angular momentum $\ell^*$ takes the form 
\begin{equation}
\ell^*=\ell_{2b}+\frac{d-3}{2},
\label{lstar}
\end{equation} 
where $\ell_{2b}$ is the relative orbital angular momentum between the two particles. 

All along the squeezing process, the momentum $\ell_{2b}$ is a good quantum number, which 
for $d=3$ becomes the usual orbital angular momentum. Also, when confining from three to two dimensions
by means of an external squeezing potential, it is known that the angular momentum 
projection quantum number, $m$, is conserved \cite{gar19b}, and, furthermore, $m$ is the 
angular momentum in the two-dimension limit \cite{nie01}. Therefore, for $d=2$ we have 
$m=\ell_{2b}$, which, due to the conservation of $m$, implies that the same equality holds 
for the initial non-squeezed 3D-state as well.

Le us assume $V(r)$ is a spherical square-well potential, that is
\begin{eqnarray} \label{e410}
    V(r) = \begin{cases}  - V_0,& \ r < r_0 \\
     \ 0 & \ r > r_0 
    \end{cases},
\end{eqnarray}
where $V_0$ ($>0$) is the depth and $r_0$ the radius. 

After introducing the
dimensionless variable $x=r/r_0$, the equation \emph{inside} the box,
$r<r_0 \; (x<1)$, then reads
\begin{align}
  \left[ - \frac{\partial^2}{\partial x^2} + \frac{\ell^*(\ell^*+1)}{x^2}
    - k^2\right] R_{in}(x)=0,
    \label{e420}
\end{align}
and similarly  outside the box, $r>r_0 \; (x>1)$, 
\begin{align}
  \left[ - \frac{\partial^2}{\partial x^2} + \frac{\ell^*(\ell^*+1)}{x^2}
    - \kappa^2\right] R_{out}(x)=0,
    \label{e430}
\end{align}
where the wave numbers are given by
\begin{eqnarray}     \label{e440}
  k = \sqrt{\frac{2\mu r_0^2(E + V_0)}{\hbar^2}} \; , \;
  \kappa = \sqrt{\frac{2\mu r_0^2 E}{\hbar^2}}.
\end{eqnarray}

Note that $\kappa$ is imaginary for bound states, but real and positive for continuum
states.

The solution inside the box, $R_{in}$, is simply given by  
\begin{eqnarray}     \label{e450}
  R_{in}(x) = A_k x j_{\ell^*}(kx)  \; ,
\end{eqnarray}
where $j_{\ell^*}$ is the spherical Bessel function of first kind and
$A_k$ is a normalization constant.  The other Bessel function, $\eta_{\ell^*}$, also a solution 
to Eq.(\ref{e420}), diverges at $x=0$ and is therefore
removed from Eq.(\ref{e450}), since the physical solution must
satisfy that $R_{in}(x)
\rightarrow 0$ for $x \rightarrow 0$. Note that $\ell^*$ is in general 
non-integer, and even negative for $s$-waves ($\ell_{2b}=0$) when $d<3$ ($\ell^*=(d-3)/2$).

\subsection{Bound states}

For a bound state ($E<0$ and $\kappa=i|\kappa|$), the solution outside the box, $R_{out}(x)$, 
takes the form
\begin{eqnarray}     \label{e460}
  R_{out}(x) = B_{\kappa} x h^{(+)}_{\ell^*}(\kappa x)  \; ,
\end{eqnarray}
where $h^{(+)}_{\ell^*}$ is the spherical Hankel function of first kind
and $B_{\kappa}$ is a normalization constant.  The other Hankel
function solution to Eq.(\ref{e430}), $h^{(-)}_{\ell^*}(\kappa x)$, is removed, since the bound state
solution must fall off exponentially at large distance.

Matching the logarithmic derivatives of Eqs.(\ref{e450}) and
(\ref{e460}) at the box radius, $x=1$, we obtain the eigenvalue equation
for bound states. The normalization factors disappear and we find, 
by use of the two identities Eqs.(\ref{A480}) and
(\ref{A490}), that
\begin{eqnarray}     \label{e470}
  1 + \ell^* - k \frac{j_{\ell^*+1}(k)}{j_{\ell^*}(k)} = 1 + \ell^* -
  \kappa \frac{h^{(+)}_{\ell^*+1}(\kappa) }{h^{(+)}_{\ell^*}(\kappa) } \;,
\end{eqnarray}
where the left and right hand sides in Eq.(\ref{e470}) are the logarithmic
derivatives of the solutions inside and outside the box, respectively.

Eq.(\ref{e470}) is an equation relating the square-well parameter,
$S_0$, and $\kappa$ in Eq.(\ref{e440}) through
\begin{eqnarray}     \label{e500}
 S_0^2 \equiv \frac{2 \mu r_0^2 V_0 }{\hbar^2} \;,\; k^2 = S_0^2 + \kappa^2 \;.
\end{eqnarray}
In other words, the bound state energy, $E$, in units of $\hbar^2/(\mu r_0^2)$, is 
a unique function of the square-well parameter, $S_0$.  The oscillatory behavior 
of the Bessel functions provides more and more discrete energy solutions with
increasing $S_0$.  All this is in complete analogy to the usual
square-well relative two-body problem, but now depending on $\ell^*$, and therefore on $d$.

\subsection{Efimov condition}

Let us consider now a moderate square-well parameter, $S_0$, which is
too small to support any bound state for $d=3$.  However, if the system
is bound in two dimensions, we then know that, when decreasing
$d$ from $3$ to $2$, a bound state of zero energy
must appear at some value, $d=d_E$. This always happens for $s$-waves ($\ell_{2b}=0$), since for $d=2$ 
a bound state is always present, even for an infinitesimal attraction. We label this dimension,
$d_{E}$, because if an additional particle is added, the resulting three-body 
system would exhibit Efimov properties, if at least two of the two-body subsystems,
in a relative $s$-wave,
are bound by precisely zero energy. In particular, if dealing with identical
particles, this critical value, $d_E$, is therefore the non-integer dimension
where the Efimov effect occurs on the
path towards two dimensions \cite{gar20}.  

To find an expression for
$d_E$, we take the limit $E\rightarrow 0^-$ in the eigenvalue
equation, Eq.(\ref{e470}).  This means that $\kappa \rightarrow 0$,
and $k\rightarrow \sqrt{2\mu r_0^2V_0/\hbar^2} = S_0$.  Using the
low-energy expansion of the Bessel-functions, Eq.(\ref{A500}), we
easily get:
\begin{eqnarray}     \label{e510}
\kappa \frac{h^{(+)}_{\ell^*+1}(\kappa) }{h^{(+)}_{\ell^*}(\kappa) } \rightarrow 2\ell^* +1\;.
\end{eqnarray}

The eigenvalue equation, Eq.(\ref{e470}), reduces then to
\begin{eqnarray}     \label{e520}
  S_0 \frac{j_{\ell^*+1}(S_0)}{j_{\ell^*}(S_0)} = 2\ell^* +1 = 2\ell_{2b}+d_E - 2 \;,
\end{eqnarray}
which is the condition for a zero-energy solution, and therefore 
gives $d=d_E$ as a function of the square-well parameter, $S_0$. We emphasize
that the index, $\ell^*$, on the Bessel function depends on $d$
as expressed in Eq.(\ref{lstar}). 

It is important to note here that the philosophy of studying scattering properties through an
extension of the angular momentum to non-integer values, $\ell^*$, is akin to the concept of Regge 
poles, where the angular momentum is generalized to a continuum, and even complex, variable, and the 
analytical behavior of the $S$-matrix is subsequently studied. In fact, Eq.(\ref{e520}) can be simplified
into:
\begin{equation}
J_{\ell^*-\frac{1}{2}}(S_0)=0,
\label{reg}
\end{equation}
which, as shown in \cite{bol62}, determines the zero-energy position of the Regge poles
for the square-well potential. Therefore, the critical dimension, $d_E$, for a given $S_0$ value 
corresponds to the zero-energy Regge pole of the potential.

\subsection{Scattering length}

The zero-energy condition is closely related to the scattering length,
which also must depend on the dimension $d$.  We derive an expression by first
noting that the zero-energy reduced radial wave function must by
definition have the form \cite{nie01}
\begin{eqnarray} \label{e530}
  R_{out}(x) \propto x^{\ell^*+1} -
  \frac{(a_d^{(\ell^*)}/r_0)^{2\ell^*+1}}{x^{\ell^*}},
\end{eqnarray}
expressed in terms of the scattering length, $a_d^{(\ell^*)}$ divided by
the length unit, $r_0$.  We used that the reduced, $R$, and total,
$\psi$, radial wave functions are related by $R(x)=x^{(d-1)/2}\psi(x)$.

Matching the logarithmic derivatives of Eqs.(\ref{e450}) and
(\ref{e530}) at $x=1$, we get
\begin{eqnarray} \label{e540}
  1 + \ell^* - S_0 \frac{j_{\ell^*+1}(S_0)}{j_{\ell^*}(S_0)} = 
    \frac{1 + \ell^* + \ell^* \big(\frac{a_d^{(\ell^*)}}{r_0}\big)^{2\ell^*+1}}
  {1 - \big(\frac{a_d^{(\ell^*)}}{r_0}\big)^{2\ell^*+1}} \; ,
\end{eqnarray}
which immediately leads to
\begin{eqnarray} \label{e550}
  \frac{a_d^{(\ell^*)}}{r_0} =
\Big({1 - \frac{(2\ell^*+1) j_{\ell^*}(S_0)}{S_0j_{\ell^*+1}(S_0)}}\Big)^{-1/(2\ell^*+1)} \;.
\end{eqnarray}
The scattering length depends on $S_0$ and $\ell^*$, or $S_0$, $d$ and
$\ell_{2b}$, in units of the square-well radius.

We see from Eq.(\ref{e520}) that this scattering length,
$a_d^{(\ell^*)}$, is infinitely large exactly when $d=d_E$.  Another
consistency check is for $d=3$ and $\ell_{2b}=0$ ($\ell^* = 0$), where
Eq.(\ref{e550}) reduces to the well-known expression:
\begin{eqnarray} \label{e560}
  \frac{a_{d=3}^{(\ell^*=0)}}{r_0} = 1- \frac{\tan(S_0)}{S_0} \;.
\end{eqnarray}

\subsection{Phase-shifts}

Let us consider now the case of continuum states ($E>0$). The large-distance
solution in Eq.(\ref{e460}) should be replaced by
\begin{eqnarray}     \label{e600}
  R_{out}(x) = x (\cos\delta_{\ell^*} j_{\ell^*}(\kappa x) -
  \sin\delta_{\ell^*}  \eta_{\ell^*}(\kappa x) )    \; ,
\end{eqnarray}
where $\eta_{\ell^*}$ is the irregular Bessel function, and
$\delta_{\ell^*}$ is the phase-shift corresponding to this
continuum state of generalized angular momentum $\ell^*$ and energy
$E$.

The solution of $\delta_{\ell^*}$ as function of $E$ and $\ell^*$ is found
by matching the logarithmic derivatives of Eqs.(\ref{e450}) and
(\ref{e600}), that is
\begin{eqnarray} \label{e610}
  \frac{kj^{\prime}_{\ell^*}(k)}{\kappa j_{\ell^*}(k)} =  
  \frac{\cot\delta_{\ell^*} j^{\prime}_{\ell^*}(\kappa) -  \eta^{\prime}_{\ell^*}(\kappa)}
       {\cot\delta_{\ell^*} j_{\ell^*}(\kappa) -  \eta_{\ell^*}(\kappa)} \; ,
\end{eqnarray}
where the prime denotes derivative with respect to the
full argument of the function.  The phase-shift can then be calculated from
Eq.(\ref{e610}) to give
\begin{eqnarray} \label{e620}
  \cot\delta_{\ell^*}  = \frac{\eta_{\ell^*}(\kappa)}{j_{\ell^*}(\kappa)} 
  \frac{\kappa \frac{\eta_{\ell^*+1}(\kappa)}{\eta_{\ell^*}(\kappa)} -
    k \frac{j_{\ell^*+1}(k)}{j_{\ell^*}(k)} }
     {\kappa \frac{j_{\ell^*+1}(\kappa)}{j_{\ell^*}(\kappa)} -
    k \frac{j_{\ell^*+1}(k)}{j_{\ell^*}(k)}}  \; ,
\end{eqnarray}
where we used the derivative expressions for both $j_{\ell^*}$ and $\eta_{\ell^*}$
from Eqs.(\ref{A480}) and (\ref{A490}).

The low-energy limits, $\kappa \rightarrow 0$, $k \rightarrow S_0$, are
then to leading order obtained from Eq.(\ref{A500}), 
\begin{eqnarray} \label{e630}
  \cot\delta_{\ell^*} &\rightarrow& \frac{-1}{(2\ell^*+1)\kappa^{2\ell^*+1}}  \\ \nonumber
 \times \bigg(\frac{\Gamma(2\ell^*+2)}{2^{\ell^*}\Gamma(\ell^*+1)}\bigg)^2 
  &\times& \bigg(1-\frac{(2\ell^*+1)j_{\ell^*}(S_0)}{S_0j_{\ell^*+1}(S_0)} \bigg) \;,
\end{eqnarray}
where $\Gamma$ is the usual Gamma function. 

Note that for $d=d_E$, making use of 
Eq.(\ref{e520}) in the last bracket of Eq.(\ref{e630}), we get
that $\cot\delta_{\ell^*}=0$, or, in other words, $\delta_{\ell^*}=\pi/2$.

Assuming that the low-energy limit of the phase-shift is proportional
to $\kappa^{2\ell^*+1}$, we can express this limit in terms of the
scattering length in Eq.(\ref{e550}), that is
\begin{eqnarray} \label{e640}
  \delta_{\ell^*} \rightarrow
  - (2\ell^*+1) \frac{2^{2\ell^*}\Gamma^2(\ell^*+1)}{\Gamma^2(2\ell^*+2)}
  \bigg(2\mu E [a_d^{(\ell^*)}]^2 /\hbar^2\bigg)^{\frac{2\ell^*+1}{2}} \;.
\end{eqnarray}

\subsection{Cross sections}
\label{crse}

As mentioned several times, the $d$-method has been introduced as a tool that simplifies the description of systems 
squeezed due to the 
presence of an external field. It is obvious that any observable will be measured in the squeezed 3D space,
and therefore, in order to compare the calculations with whatever available experimental data, it is necessary
to translate the observable from the $d$-dimensional space into the 3D space. 

This is in particular necessary when computing cross sections. It is not obvious that the phase shifts derived in the
previous subsection can be directly employed to get the cross sections simply by using them in the usual 3D cross section expressions in
terms of the phase shifts. 

To understand how to proceed, we first interpret the $d$-dimensional wave function as
described in \cite{gar19a}, that is, as a wave function in the ordinary 3D space where the relative coordinate
in the $d$-space, $\bm{r}$,
is now understood as a 3D vector, denoted as $\tilde{\bm{r}}$, but where the third coordinate (the $z$-axis is chosen 
along the squeezing direction) is squeezed by means of a scale parameter $s$. In other words, the $d$-coordinate, $\bm{r}$, is then understood as a 3D-coordinate as:
\begin{equation}
\bm{r} \rightarrow \tilde{\bm{r}}=(x,y,\tilde{z})=(x,y,\frac{z}{s}),
\end{equation}
where $x$, $y$, and $z$ are the usual Cartesian coordinates in 3D of the actual relative coordinate, $r$, in the 3D-squeezed
space. In this way:
\begin{equation}
\tilde{r}^2=x^2+y^2+\frac{z^2}{s^2}=r_\perp^2+\frac{z^2}{s^2}=r^2\left( \sin^2\theta+\frac{\cos^2\theta}{s^2} \right),
\label{eq5}
\end{equation}
where $\theta$ is the usual polar angle.

In the same way, the direction of the 3D-coordinate, $\tilde{\bm{r}}$, is given by the polar and azimuthal angles $\tilde{\theta}=\arctan(r_\perp/\tilde{z})$ and $\tilde{\varphi}=\arctan(y/x)$. They can be easily related to the polar and 
azimuthal angles of the actual relative coordinate, $\theta=\arctan(r_\perp/z)$ and $\varphi=\arctan(y/x)$,
leading to:
\begin{equation}
\tan \theta=\frac{1}{s}\tan{\tilde{\theta}}, \hspace*{5mm} \varphi=\tilde{\varphi},
\label{tilth}
\end{equation}
which reflects the fact that the squeezing is taken along the $z$-axis, and in the case of large squeezing
(very small $s$) only values of $\theta$ very close to $\theta=\pi/2$ are allowed, as it corresponds to a system 
moving in the $xy$-plane.

To determine the value of the scale parameter $s$ we use the approximate expression derived in \cite{gar20}:
\begin{equation}
\frac{1}{s^2}=\left[1+\left( \frac{(3-d)(d-1)}{2(d-2)} \right)^2 \right]^{1/2},
\end{equation}
which is based on the assumption of harmonic oscillator particle-particle interactions. For identical particles, only the harmonic frequency of the interaction enters and it can be adjusted to give the same $d=2$ binding energy as the general short-range interaction. Numerical tests show that this is a  good approximation \cite{gar20}.

However, for continuum states this interpretation of the radial coordinates is still not enough. Even for very small
values of the scale parameter $s$ (which imply a large squeezing along the $z$-direction) the asymptotic wave
function (\ref{e600}) is still oscillating, and not vanishing, for sufficiently large values of $z$. Therefore, it
is necessary to introduce an additional factor imposing the confinement along the squeezing direction.
In particular, assuming a harmonic oscillator squeezing, we write the asymptotic form as:
\begin{equation}
R(\tilde{r}) \stackrel{r\rightarrow \infty}{\longrightarrow}  \left[ \tilde{r}  
\left(\cos\delta_{\ell^*} j_{\ell^*}(k\tilde{r})-\sin\delta_{\ell^*} \eta_{\ell^*}(k\tilde{r}) \right) \right]
e^{-\frac{z^2}{2 b_{ho}^2}},
\label{eq3b}
\end{equation}
where $z=r\cos\theta$ and $b_{ho}$ is the harmonic oscillator length associated to the squeezing potential. According to our estimates \cite{gar20}, the connection between $d$ and $b_{ho}$ is given by:
\begin{equation}
b_{ho}=\sqrt{\frac{2(d-2)}{(d-1)(3-d)} },
\end{equation}
which is given in units of the range of the particle-particle interaction.

The idea now is to interpret the $d$-wave function as an ordinary wave function in three dimensions whose 
radial part depends on $\tilde{r}$, and and whose angular part is an ordinary spherical harmonic depending
on $\tilde{\theta}$ and $\tilde{\varphi}=\varphi$. Remembering that in 3D, as discussed below Eq.(\ref{lstar}), 
we must have that $m=\ell_{2b}$ we can then write:
\begin{equation}
\frac{R(\tilde{r})}{\tilde{r}^{\frac{d-1}{2}}} Y_{\ell_{2b}\ell_{2b}}(\tilde{\theta},\varphi)
 = \sum_{\ell m} \frac{u_\ell(r)}{r} Y_{\ell m}(\theta,\varphi),
\label{twf}
\end{equation}
such that we expand the wave function in terms of the spherical harmonics expressed as function of the
polar and azimuthal angles of the 3D relative coordinate.

From the equation above we can extract the projected radial wave functions as:
\begin{equation}
\frac{u_\ell(r)}{r}= \int d\Omega \frac{R(\tilde{r})}{\tilde{r}^{\frac{d-1}{2}}} 
 Y_{\ell_{2b}\ell_{2b}}(\tilde{\theta},\varphi) Y_{\ell m}^*(\theta,\varphi),
\end{equation}
which, after integrating over $\varphi$ (leading to $\delta_{m,\ell_{2b}}$) and getting rid of the constants factors, becomes:
\begin{equation}
u_\ell(r) \propto r \int_0^\pi d\theta \sin\theta \frac{R(\tilde{r})}{\tilde{r}^{\frac{d-1}{2}}}  
P_{\ell_{2b}}^{(\ell_{2b})}(\cos\tilde{\theta}) P_\ell^{(\ell_{2b})}(\cos\theta) 
\label{eq9}
\end{equation}
whose asymptotic behavior can be computed by introducing Eq.(\ref{eq3b}).
For a given $r$, the integrand in the equation above is just a function of $\theta$ 
through Eqs.(\ref{eq5}) and (\ref{tilth}). 

By fitting the $u_\ell$ functions to the expected asymptotic behavior, 
\begin{equation}
u_\ell(r) \stackrel{r\rightarrow \infty}{\longrightarrow} 
r  \left(\cos\delta_\ell j_{\ell}(kr)-\sin\delta_\ell \eta_{\ell}(kr) \right)
\label{eq12}
\end{equation}
we can extract the phase shift for each partial wave $\ell$.

However, it is important to keep in mind that Eq.(\ref{eq12}) gives the asymptotic
behavior of the continuum wave functions in the ordinary, non-squeezed, 3D space. 
The effect of the confining external potential is not present in Eq.(\ref{eq12}).
In fact, in the case of no interaction between the particles, the $d$-phase shift,
$\delta_{\ell^*}$, obtained from the asymptotic behavior (\ref{e600}) is trivially
equal to zero, and the radial wave function is simply given by $x j_{\ell^*}(kx)$.
When introducing this function into Eq.(\ref{eq9}), and making use of Eq.(\ref{eq12}),
we can easily observe that the corresponding computed 3D phase shift is not zero. This non-zero
phase shift, which we denote as $\delta_{\mbox{\scriptsize free}}$, reflects the effect
of the confining external potential.

By definition, the phase shift is just the shift of the asymptotic wave function when compared
to the free wave function. It is then clear that the phase shift in the confined 3D space,
$\delta_{\mbox{\scriptsize conf}}^{(\ell)}$, is then given by:
\begin{equation}
\delta_{\mbox{\scriptsize conf}}^{(\ell)}=\delta_\ell - \delta_{\mbox{\scriptsize free}},
\label{eq32}
\end{equation} 
where $\delta_\ell$ and $\delta_{\mbox{\scriptsize free}}$ are then the phase shifts obtained from 
Eqs.(\ref{eq9}) and (\ref{eq12}) for interacting and free particles, respectively. 

We shall later show that $\delta_{\mbox{\scriptsize conf}}^{(\ell)}$ is actually independent of $\ell$, which
permits to associate a single phase shift $\delta_{\mbox{\scriptsize conf}}\equiv 
\delta_{\mbox{\scriptsize conf}}^{(\ell)}$, to the two-body continuum state in the relative $\ell_{2b}$-wave 
in the squeezed 3D space. Furthermore, we shall
also show the nice result that $\delta_{\ell^*}=\delta_{\mbox{\scriptsize conf}}$, which in fact means that
the phase shifts obtained with the $d$-method are actually the phase shifts in the
squeezed 3D space, and therefore they can be directly used to compute the cross section
in the 3D space, which is given by the simple well-known formula:
\begin{equation}
\sigma=\frac{4\pi}{k^2} (2\ell_{2b}+1) \sin^2\delta_{\ell^*}.
\end{equation}

\section{Results}
\label{sectIII}

In this section we show the results for the key quantities analyzed
in the previous section for two-body continuum states, i.e., 
critical dimension, scattering length, and phase-shifts. The derivations
have been carried out assuming a square-well potential, and therefore the
derived formulae, which of course depend on the dimension $d$, will depend 
as well on the square-well parameter, $S_0$. This dependence has been obtained
in a very specific model, and provides results that, in principle, are not
universal. However, as we shall show, it is possible to
extract a universal behavior, such that the square-well results previously
derived can be exported to any short-range potential.

\subsection{Critical dimension}

\begin{figure}[ht]
    \centering
    \includegraphics[scale=0.5]{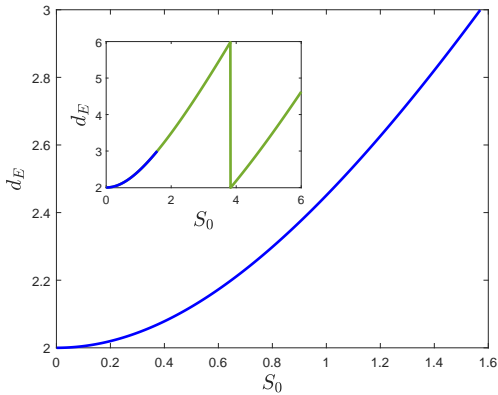}
    \caption{The critical dimension, $d_E$, for $s$-waves, as a function of the
      potential parameter, $S_0$, defined in Eq.(\ref{e500}).}
    \label{dc2}
\end{figure}

The critical dimension, $d_E$, is obtained by solving Eq.(\ref{e520})
as function of $S_0$. This dimension always exists for relative $s$-waves
($\ell_{2b}=0$), which is particularly relevant, since this is the dimension
for which the Efimov effect arises in the three-body case.

The resulting dependence, shown in
Fig.~\ref{dc2} for $\ell_{2b}=0$, is a smoothly increasing function of $S_0$,
changing from $d_E=2$ at $S_0=0$, to $d_E=3$ for $S_0=\pi/2$. 
This reflects the fact that an infinitely small attractive potential is, for relative $s$-waves, 
sufficient to bind for $d=2$, but a finite potential is necessary to support bound states
as the dimension approaches $d=3$. The value of $S_0 = \pi/2$ for $d=3$
is the well-known critical size for binding in a square-well. Note also
that, as shown in Eq.(\ref{reg}), for $d=3$ the values of the critical
dimension, $d_E$, appear
for the $S_0$ values matching with the zeros of $J_{-1/2}(S_0)$, which
are given by $S_0=(2n+1)\pi/2$, with $n=0,1,2,\cdots$.

Increasing the attraction of the potential still smoothly increases
$d_E$ above $3$, as shown in the inset of Fig.~\ref{dc2}.  This
behavior is abruptly broken by a sudden downwards jump at $S_0 = 3.83171$,
where a second bound state appears at zero energy. The following
increase is the beginning of repeating this behavior reflecting the
discrete increase of bound states. In fact, again, we can from
Eq.(\ref{reg}) see that for $d=2$ the $d_E$ values for which new
bound states appear are determined by the zeros of $J_{-1}(S_0)=-J_1(S_0)$,
whose first three values are $S_0=3.83171, 7.01559, 10.17347$, and which give 
the position of the jumps in the curve shown in the inset of Fig.~\ref{dc2}.

The square-well parameter is by definition not a universal quantity
that could be used for other short-range potentials. Instead, in order to
pursue such a universality, we turn to the scattering length for $d=3$, which is 
a unique characteristic for any short-range potential, as well as definable, measurable, 
and reflecting a gross property independent of potential details.
It determines rather accurately all low-energy scattering properties, and
as such is a universal quantity.

Although in principle possible as well for relative angular momenta larger than zero, universality
features are specially relevant for $s$-waves, for which the wave function can more easily
extend beyond the range of the interaction, becoming then less sensitive to the details
of the potential. As shown in Eq.(\ref{e640}), for small energy
the phase shifts behave as $E^{(2\ell_{2b}+d-2)/2}$, meaning that the $s$-wave scattering 
clearly dominates in the low-energy limit, and for this reason we shall focus here on universal behavior in
the $\ell_{2b}=0$ case (an exception could be the case, not considered here, of two identical fermions, 
for which a relative $s$-wave is not possible \cite{bra06}).

\begin{figure}[t]
\includegraphics[scale=0.6]{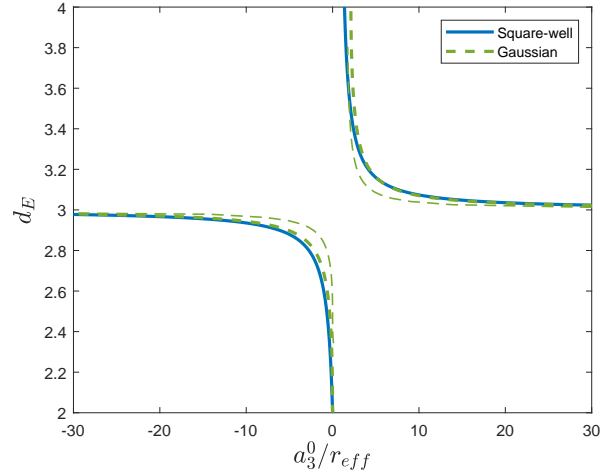}
    \caption{The critical dimension, $d_E$, as a function of the scattering length, 
    $a^{(\ell_{2b}=0)}_{d=3}/r_{\mbox{\scriptsize eff}}$, for the square-well (blue thick-solid) and the 
    Gaussian potentials (green thick-dashed). The dimension $d_E$ corresponds to the appearance of
    the first bound state. The green thin-dashed curve shows the $d_E$-values corresponding
    to the appearance of the second bound state with the Gaussian potential.}
    \label{dc3}
\end{figure}

In Fig.~\ref{dc3} we show the critical dimension, $d_E$, as a function
of $a^{(\ell_{2b}=0)}_{d=3}$ for both square-well and Gaussian potentials.
Different length units have been considered, and we have found that 
clearly the length scale with the effective range, $r_{\mbox{\scriptsize eff}}$, 
provides the most satisfactory universal behavior. 
This is in accordance with the findings for three-body systems \cite{nai14}. The results for the 
two potentials are pretty close to each other, and we therefore have,
to a very large extent, the desired, accurate universal function,
$d_E$, as function of scattering length in units of the effective
radius, both obtained in the ordinary three dimensions. This is what shown by the two thick curves 
in the figure, solid (blue) and dashed (green), which have been obtained with potential strengths such 
that $d_E$ corresponds to the appearance of the first bound state, i.e., to potential strengths 
unable to bind the system in three dimensions. 

In principle one could still increase $S_0$ such that a second
bound state appears (as indicated by the jump in the curve shown in the inset of Fig.~\ref{dc2}).
As exemplified in previous works, the depth of the two-body potential could be an important factor 
for universality \cite{nai14,mes19}. Although this interesting relationship goes
beyond the scope of this work, we show in Fig.~\ref{dc3}, as an indication, the
curve (thin-dashed) obtained for the critical dimension $d_E$ corresponding to the 
appearance of the second bound state with the Gaussian potential. As we can see, this curve does 
not follow the universal curve obtained when $d_E$ indicates the appearance of the
first bound state.

The scattering length and effective range ratio for $\ell_{2b}=0$ and $d=3$ is therefore 
adopted as the easy accessible universal quantity.  This has the advantage of 
requiring only one adjustment, the $a_{d=3}^{(\ell_{2b}=0)}/r_{\mbox{\scriptsize eff}}$ ratio,
when two potentials should be compared.  Another possibility could perhaps be to use the
same $a^{(\ell^*)}_d$ for different potentials as defined in
Eq.(\ref{e550}), or may be the same ratio between $a^{(\ell^*)}_d$ and
the corresponding $d$-dependent effective range. However, this would require calculation 
with different potentials for any $d$.  The simplicity of the method would be
lost, and we therefore we maintain the simple procedure.

\subsection{Scattering length}

In any case, even if not used in the universal curve shown in Fig.~\ref{dc3}, a given
two-body potential, has associated a different scattering length for different values
of the dimension $d$, as shown in Eq.(\ref{e550}) for the case of the square-well
potential.

\begin{figure}[t]
    \includegraphics[scale=0.5]{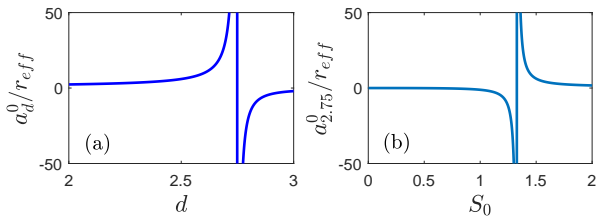}
    \caption{(a) Scattering length, $a^{(\ell_{2b}=0)}_d$, as function of $d$
      for $S_0 \simeq 4/3$, such that $d_E=2.75$. (b) Scattering length as function of $S_0$
      for $d=2.75$.}
    \label{scatlenplot}
\end{figure}

This $d$-dependent scattering length is a function of $S_0$, as shown in
Fig.~\ref{scatlenplot} for the square-well potential and $\ell_{2b}=0$. The left and right parts 
show the dependence on $d$ for a given $S_0$ ($S_0\simeq 4/3$ such that $d_E=2.75$), and on $S_0$ for a given 
$d$ ($d=2.75$), respectively.  If desired $S_0$ can, through Fig.~\ref{dc2}, be substituted by $a^{(\ell_{2b}=0)}_{d=3}$.
The qualitative features are precisely the same.

The most pronounced feature is that the scattering length diverges at
the critical dimension, $d_E$, see Fig.~\ref{scatlenplot}a. 
For $S_0\simeq 4/3$ we can see from Fig.~\ref{dc2} that $d_E\approx 2.75$,
which is the $d$-value where the divergence of the scattering length shows up. This
demonstrates the existence of a zero-energy bound state for this
dimension.  The picture would be very similar for other values of $S_0$,
where the zero-energy would appear at another $d$.  For illustration,
we show in Fig.~\ref{scatlenplot}b the other dependence of the
scattering length as a function of $S_0$ for given $d$ ($d=2.75$). 
Again we observe a divergence of $a^{(\ell_{2b}=0)}_d$ for a specific $S_0$, $S_0\approx 4/3$, 
which agrees with the value observed in Fig.~\ref{dc2} for $d_E=2.75$, and defines the parameter 
set producing a zero-energy bound state.

\subsection{Phase-shifts}

\begin{figure*}[ht]
    \includegraphics[width=0.75\textwidth]{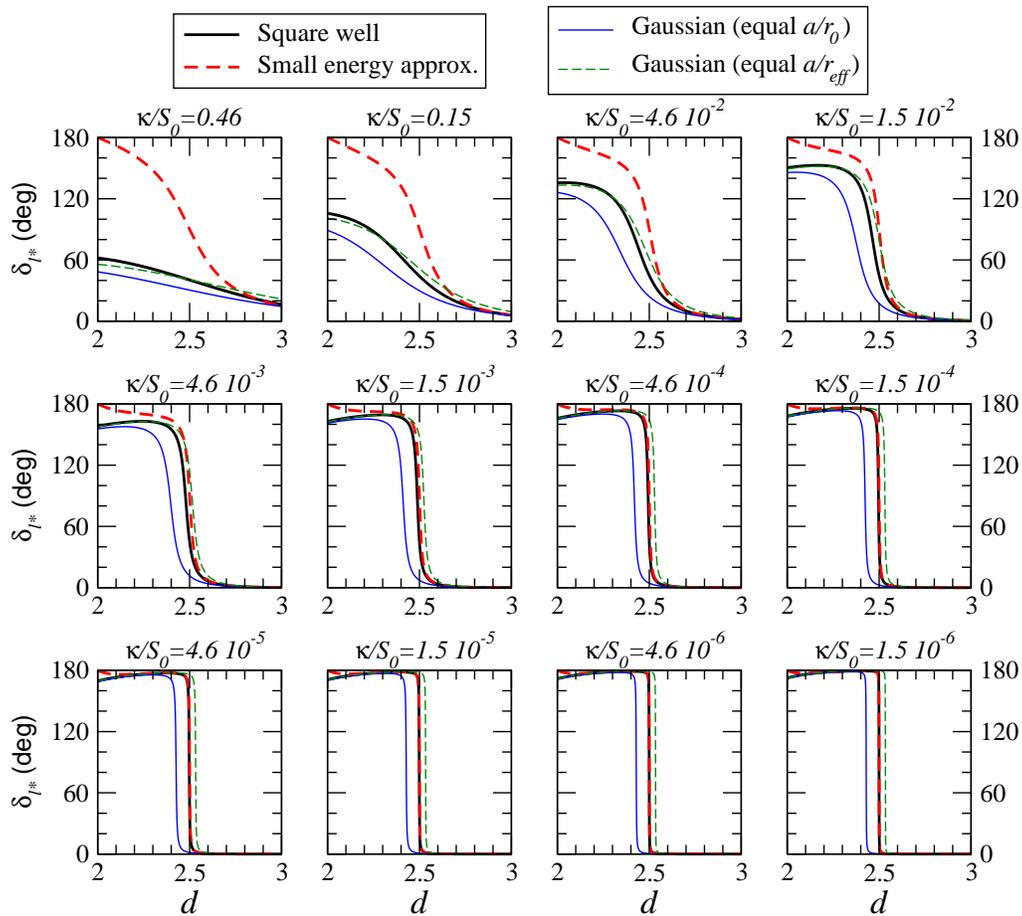}
     \caption{Phase shifts, $\delta_{\ell^*}$, for $\ell_{2b}=0$ as function of $d$ for different values of
       $\kappa/S_0$.  The thick-solid-black curves are square-well results from
       Eq.(\ref{e640}), the thick-dashed-red curves are the square-well
       low-energy approximation from Eq.(\ref{e630}), the thin-solid-blue curves
       are for a Gaussian interaction comparing results for equal
       $a_{d=3}^{(\ell_{2b}=0)}/r_0$ (scattering length divided by the Gaussian or square-well
       range) in 3D, and the thin-dashed-green curves are for a Gaussian interaction
       comparing results for equal $a_{d=3}^{(\ell_{2b}=0)}/r_{\mbox{\scriptsize eff}}$ (scattering length
       divided by effective range) in 3D. The potential parameters produce $d_E=2.50$ for the square-well interaction. }
    \label{phasetwo}
\end{figure*}

The phase-shifts contain the crucial information about the continuum states.
They provide the information about the asymptotic behavior of the wave function, and 
they are the key ingredient to obtain an important observable like the
cross sections. 

In Fig. \ref{phasetwo} we show the dependence of the $s$-wave phase shifts on 
the dimension $d$ and on the energy (through $\kappa$, Eq.(\ref{e440})). 
Each sub-figure compares four computed phase-shifts as function of the $d$ for given
values of $\kappa/S_0$. 

The first two curves on each panel, thick-solid-black and thick-dashed-red, are the analytic square-well results 
in Eq.(\ref{e620}) and the related low-energy approximation in Eq.(\ref{e640}), 
respectively. Not
surprisingly, we find the approximation improves with decreasing
energy (decreasing $\kappa$), but apparently also with increasing $d$.  The mathematical
reason can be found in the size of the $d$-dependent expansion
parameter leading to Eq.(\ref{e640}). The parameters for the square-well potential
have been chosen such that $d_E=2.50$. The consequence is that for this value
of $d$, as mentioned below Eq.(\ref{e630}), the low energy curves (thick-dashed-red)
always take the value $\delta_{\ell^*}=\pi/2$.

For $d=2$ the low-energy phase-shifts are always $\pi$ corresponding
to a bound state.  The exact model results are also close to $\pi$ for
$d \approx 2$ and sufficiently small energies.  The validity range of
the low-energy expansion seem to be energies less than $\kappa/S_0
\lesssim 0.001$.  For all these lower energies we observe a very sharp
transition around $d_E$ from phase-shifts of $\pi$ to zero moving
towards larger $d$-values.  This behavior is connected to the general
property of at least one bound state for any attraction for $d=2$ and
in our case for $d<d_E$.  The phase shift is passing $\pi/2$ and a
bound state of zero energy appears for $d=d_E$.  For $d>d_E$ there is
no bound state, but a negative centrifugal barrier leading to
phase-shifts close to zero, whereas there is at least one bound state
for $d<d_E$, perhaps also conditions corresponding to a resonance with
phase shifts $\pi$.

In Fig.~\ref{phasetwo} the results arising from the use of a Gaussian 
potential are also shown. To investigate the proper comparison between 
different potentials we first consider the square-well and the Gaussian
potentials having the same scattering length ratio to the 
interaction range, $r_0$, for $d=3$. The results for the Gaussian
case are given by the thin-solid-blue curves. When compared to the
square-well results (thick-solid-black) we see that both curves are
quantitatively very similar, although the agreement between them is not perfect.
In fact, the Gaussian potential constructed as mentioned above leads to 
a critical dimension $d_E=2.43$, which implies that the crossing 
through $\delta_{\ell^*}=\pi/2$ at low energies can be easily distinguished 
from the one of the square-well potentials, which happens at $d_E=2.50$.

However, when the Gaussian potential is constructed having the
same ratio between the scattering length and the effective range
as in the square-well potential for $d=3$, we then observe a clearly better
agreement between both cases, as seen when comparing the thin-dashed-green
and the thick-solid-black curves in Fig.~\ref{phasetwo} (the Gaussian potential 
has now $d_E=2.53$).  The universal function describing the
critical dimension as function of the scattering length divided by
the effective range is therefore also valid for continuum states.  This result is
consistent with the universal curve shown in Fig.~\ref{dc3}, also a function
of $a_{d=3}^{(\ell_{2b}=0)}/r_{\mbox{\scriptsize eff}}$. 

This is
very reassuring, because we can then compare potentials with the same
gross properties like scattering length and effective range, and claim
the computed universal relation between phase shifts and dimension. In
other words one can use the analytic square-well results and translate
to any other potential by use of the potential-independent parameters,
scattering length and effective range.

\subsection{Cross sections}

As described in Section~\ref{crse}, the $d$-dimensional wave function is interpreted as a wave function 
in the squeezed 3D space. As shown in Eq.(\ref{twf}), this wave function can be expanded in terms
of the usual spherical harmonics, which permits to extract the projected
radial wave functions $u_\ell(r)$ as shown in Eq.(\ref{eq9}).

\begin{figure*}
\centering
\includegraphics[width=0.75\textwidth]{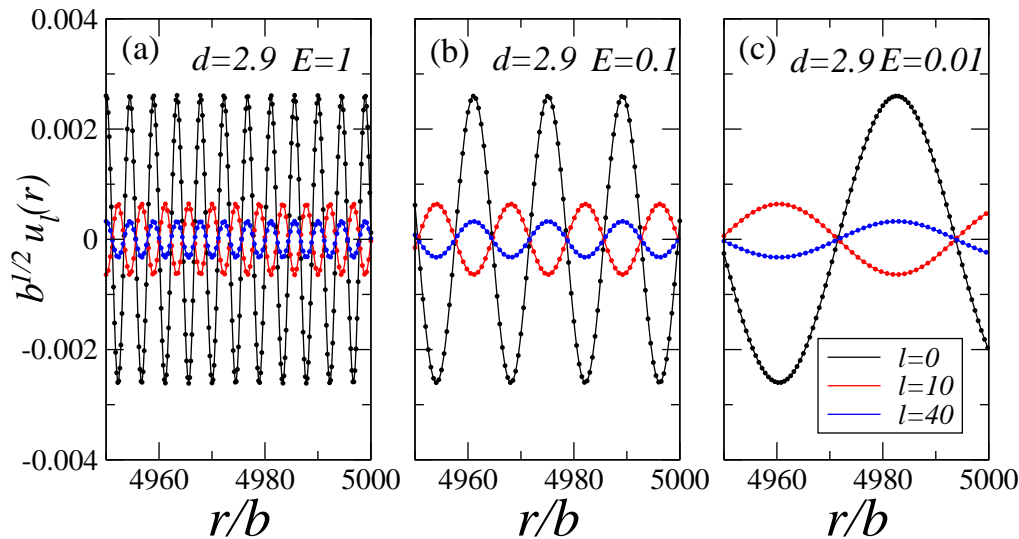}
\caption{Projected radial wave functions $u_\ell(r)$ for $\ell_{2b}=0$, $d=2.9$, and $\ell=0$ (black), 10 (red), and 40 (blue) 
and the Gaussian potential in Ref.~\cite{gar21}. The incident energy (in units of $\hbar^2/(\mu b^2)$) is equal to 1 (a), 0.1 (b), and 0.01 (c). The solid curves are the results from
Eq.(\ref{eq9}), and the dots are the asymptotic behavior given in Eq.(\ref{eq12}). The wave functions, $u_\ell(r)$, and
the radial coordinate, $r$, are multiplied by $b^{1/2}$ and divided by $b$, respectively, to make them dimensionless
($b$ is the range of the potential).}
\label{fig1}
\end{figure*}

In Fig.~\ref{fig1} we show, for $\ell_{2b}=0$, the large distance part of the $u_\ell(r)$ functions after the calculation in Eq.(\ref{eq9}) 
for $d=2.9$, $\ell=$ 0 (black), 10 (red), and 40 (blue), and three different energies, $E=1$ (a), $E=0.1$ (b), and 
$E=0.01$ (c) in units of $\hbar^2/(\mu b^2)$, where $b$ is the range
of the Gaussian two-body interaction used in the calculation. To be precise, we have used the Gaussian two-body
potential given in Ref.~\cite{gar21}, for which $d_E=2.75$. In the figure we have multiplied $u_\ell(r)$ by $b^{1/2}$ and
divided $r$ by $b$ in order to make both quantities dimensionless.

We can see that the larger $\ell$, the smaller the weight of the corresponding $u_\ell(r)$ wave function. Therefore, one 
would expect that for sufficiently high values of $\ell$, the contribution of the $u_\ell$ wave functions to the total wave 
function (\ref{twf}) can be neglected. In general, the larger the squeezing, the larger the number of $\ell$-values
contributing. In fact, for $d=3$ only the $u_{\ell=0}(r)$ function enters.

After matching the $u_\ell$ functions with the analytic asymptotic behavior in Eq.(\ref{eq12}) we can extract the $\delta_\ell$ 
phase shifts. The result of this matching is shown in the figure by the dots, which we can see lie very much on top
of the solid curves. We can also see that, as expected, the smaller the energy, the slower the oscillations
of the wave functions. An important point in the figure is that, for each energy $E$, all the wave functions, 
independently of $\ell$, have the zeros at the same $r$-values. Although not shown in the figure, this happens no matter the value chosen for the dimension $d$. Therefore, as one could expect, it is possible to assign a single phase shift, $\ell$-independent,
to the 3D wave function introduced in Eq.(\ref{twf}).

\begin{figure}
\centering
\includegraphics[width=8.cm]{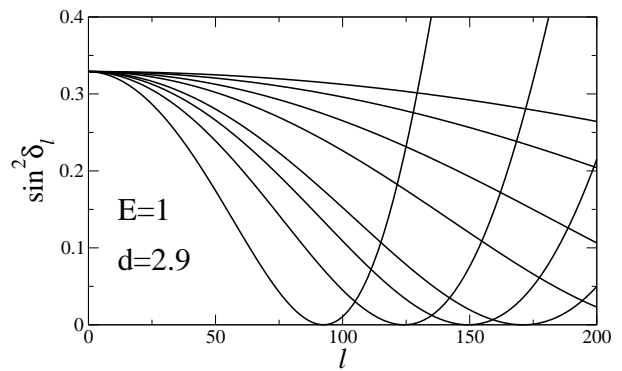}
\caption{Value of $\sin^2\delta_\ell$, as a function of the angular momentum $\ell$, for the 
$E=1$ case in Fig.~\ref{fig1}. The different curves correspond to different values of the large $r$-values used 
to extract $\delta_\ell$.
The larger the $r$-values, the closer the curve to a constant value of $\delta_\ell$.}
\label{fig2}
\end{figure}

We illustrate this same result in Fig.~\ref{fig2}, where we show $\sin^2\delta_\ell$ as a function of $\ell$ for the case 
in Fig.~\ref{fig1}a, i.e., $\ell_{2b}=0$, $d=2.9$, and $E=1$. The different curves in the figure correspond to different values of the
radial coordinate $r$ used to match the computed $u_\ell(r)$ function and the asymptotic
behavior in Eq.(\ref{eq12}). The result is that the larger the value of $r$ chosen for the matching, the closer
the curve to a horizontal line, corresponding to a constant value of $\delta_\ell$. This is due to the fact that the 
larger the value of $\ell$, the farther one has to go in the wave function in order to get the correct asymptotic behavior.
Again, the results shown in Fig.~\ref{fig2} are qualitatively the same as the ones obtained for a different value of 
$d$ and a different energy.

Finally, as mentioned in Eq.(\ref{eq32}), the phase shift, $\delta_{\mbox{\scriptsize conf}}$, in the squeezed 3D space is not given just
by $\delta_\ell$, but by the shift of the wave function for two interacting particles compared to the wave function 
corresponding to two free particles. The phase shift obtained from Eq.(\ref{eq12}) for two non-interacting particles
is what in Eq.(\ref{eq32}) is denoted by $\delta_{\mbox{\scriptsize free}}$.

\begin{figure}
\centering
\includegraphics[width=8.5cm]{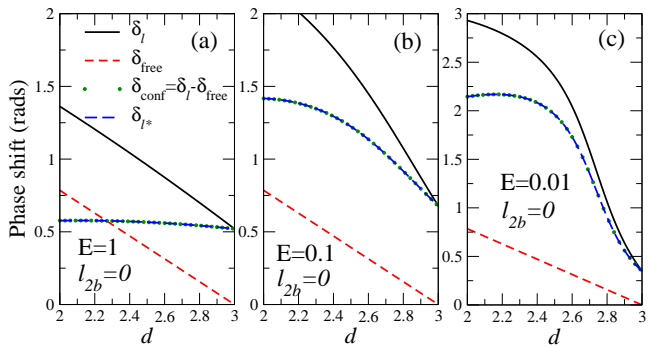}
\caption{For $\ell_{2b}=0$, values, as a function of $d$, for the phase shifts $\delta_\ell$ (solid-black), 
$\delta_{\mbox{\scriptsize free}}$ (dashed-red), $\delta_{\mbox{\scriptsize conf}}$ (dot-green),
and $\delta_{\ell^*}$ (long-dashed-blue), for the same potential and energies as in Fig.~\ref{fig1}.}
\label{fig3}
\end{figure}

In Fig.~\ref{fig3} we show $\delta_\ell$, 
$\delta_{\mbox{\scriptsize free}}$, $\delta_{\mbox{\scriptsize conf}}$,
and $\delta_{\ell^*}$ as a function of $d$ for the same Gaussian potential as in Fig.~\ref{fig1}, 
and the same three energies. The phase shifts are given by the solid-black, dashed-red, dot-green,
and long-dashed-blue curves, respectively.

Except for $d=3$, where by definition they are the same, we can see that $\delta_\ell$ and
the phase shift computed directly with the $d$-method, $\delta_{\ell^*}$, are very different.
Also, when the two particles do not interact, the phase shift, $\delta_{\mbox{\scriptsize free}}$, 
obtained from Eq.(\ref{eq12}) is clearly different from zero (except of course for $d=3$, where
the asymptotics in Eq.(\ref{eq12}) is exact). The important result is that, as shown 
by the dotted curves in the figure, the difference
between $\delta_{\ell}$ and $\delta_{\mbox{\scriptsize free}}$, that is, $\delta_{\mbox{\scriptsize conf}}$,
coincides perfectly with $\delta_{\ell^*}$, i.e., with the phase shift obtained directly
from the $d$-method.

Therefore, as anticipated in Section~\ref{crse}, the phase shifts obtained with the $d$-method are 
the phase shifts in the squeezed 3D space, and they can be used to compute the cross section
in the 3D space.

\begin{figure}
\centering
\includegraphics[width=8.5cm]{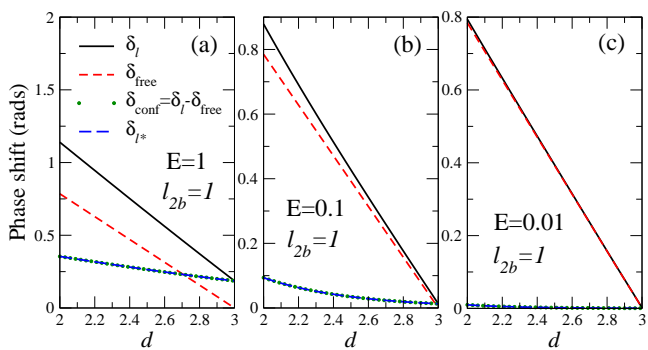}
\caption{The same as Fig.\ref{fig3} for $\ell_{2b}=1$.}
\label{fig4}
\end{figure}

The result shown in Fig.~\ref{fig3} is actually general, not restricted to the $\ell_{2b}=0$ case. As an illustration
we show in Fig.~\ref{fig4} the same as in Fig.~\ref{fig3} but for $\ell_{2b}=1$. As we can see, the
equality $\delta_{\mbox{\scriptsize conf}}=\delta_{\ell^*}$ is also perfectly obtained, even in the case of 
very small energy, Fig.~\ref{fig4}c, where $\delta_{\ell}$ and $\delta_{\mbox{\scriptsize free}}$, are 
very similar, leading to very small values of the true phase shift.

\section{Summary and conclusions}
\label{sectV}

In this work we have used the $d$-method in order
to investigate two-body continuum states in a squeezed scenario. 
The method describes the system in a non-integer dimensional space, which
can be later translated to the usual 3D space with an external confining
potential. The $d$-method is technically simpler, since the degrees of
freedom associated to the external field do not enter in the formalism.

 We focus on two-body
properties for short-range potentials in non-integer geometry by use
of a square-well potential.  The details are independent of this practical
choice of the potential.  

We first derive conditions for bound states
expressed by the square-well parameter. The depth-radius combination
has to be large enough to support at least one bound state.  We show
how the threshold value of the dimension for binding is uniquely
determined by the square-well parameter.  This is in close analogy to
three dimensions, but for smaller dimensions less attraction is
necessary, and for $d=2$ even infinitesimally small attraction
provides binding.  

The point of zero energy binding 
defines the Efimov dimension, $d_E$, since for this dimension the 
Efimov effect may become effective by adding another particle.
We have derived the equation determining the value of $d_E$ as 
a function of the strength of the square-well potential. This value
of $d_E$ is shown to correspond to infinite scattering length, which 
in turn has also been derived as a function of the same strength
for each dimension $d$. 

We have closed the theoretical part of this paper deriving the expressions
for the $d$-dependent phase shifts as a function of the potential strength.
These phase shifts have also been related to the ones corresponding to the
physical squeezed 3D space. 

The main purpose of the illustrations shown in this work is to show that,
even if the derived expressions, obtained for a square-well two-body
interaction, are not necessarily universal, it is however possible, using
the proper length and energy units, to obtain a universal behavior, independent
of the details of the potential, for the Efimov dimension $d_E$ as well
as for the phase shifts.

In particular, focusing on the dominant $s$-wave case, we have shown that the critical 
dimension $d_E$ follows a universal path when plotted as a function of 
$a_{d=3}^{(\ell=0)}/r_{\mbox{\scriptsize eff}}$, showing that the effective
range in three dimensions is the appropriate length unit. 
Given the value of $a_{d=3}^{(\ell=0)}/r_{\mbox{\scriptsize eff}}$ for any
two-body short-range potential it is then possible, by means of the universal
curve, to obtain the corresponding dimension $d_E$.

It is reassuring that potentials having the same value
 of the $a_{d=3}^{(\ell=0)}/r_{\mbox{\scriptsize eff}}$ ratio follow
 as well a pretty much universal curve for the phase shifts as a function
 of the dimension and for given values of $\kappa/S_0$. In other
 words, for the universal curve to show up, the incident momentum has
 to be taken in units of $S_0$, which contains the dependence on the strength 
 of the potential.

Finally, we have related the computed phase shifts in the $d$-method with the
ones corresponding to a squeezed 3D space. This is done by considering the
$d$-dimensional two-body wave function as an ordinary wave function in 3D, but
squeezing the coordinate along the direction of the external field after inclusion
of a scale parameter. When this is done, we have shown that, in case of
non-interacting particles, the usual asymptotic form of the continuum wave
functions in three dimensions gives rise to non-zero phase shifts,
$\delta_{\mbox{\scriptsize free}}$. This is simply reflecting the presence of the 
squeezing potential. When the interaction is introduced, the relevant phase
shift is the difference between the new computed phase shift and $\delta_{\mbox{\scriptsize free}}$.
We have shown that this difference, $\delta_{\mbox{\scriptsize conf}}$, is precisely
the same as the phase shift obtained directly from the $d$-method. This result
opens the door to using the $d$-computed phase shifts directly in order to obtain
the cross sections in a squeezed scenario.

In conclusion, we have investigated and illustrated two-body scattering
processes in an analytic schematic model.  The cross section behavior
and the insight obtained are very hard to imagine found in any other
way.  Such results are, to a large extent, universal, that is independent of the details of
the employed short-range potentials.  The translation from $d$ to
external field is necessary, available and at least a semi-accurate
description.  The perspective in our investigations is that scattering
between particles confined by deformed external fields may be useful
tools in investigations of for example structures related to Efimov
physics. Transitions between other dimensions may also be of interest.

\begin{acknowledgements}
This work has been partially supported by
the Spanish Ministry of Science, Innovation and University
MCIU/AEI/FEDER,UE (Spain) under Contract No. PGC2018-093636-B-I00.
\end{acknowledgements}

\begin{appendix}
  
\section{Bessel function properties}

The two regular spherical Bessel functions with indices $0$ and $1$
are
\begin{eqnarray}     \label{A470}
  j_0(z) = \frac{\sin z}{z} \; \; , \;
  j_1(z) = \frac{\sin z}{z} -\frac{\cos z}{z^2} \;.
\end{eqnarray}
We have two useful identities between any, $B_l$, of the regular,
$j_l$, and irregular, $\eta_l$, Bessel and Hankel, $h^{(\pm)}_{l}=\eta_l
\pm i j_l$ functions, that is
\begin{eqnarray}     \label{A480}
  z \frac{d B_l(z)}{d z} = l B_l(z) - z B_{l+1}(z) \; ,\\  \label{A490}
 (2l+1) B_l(z) = z (B_{l+1}(z) + B_{l-1}(z)) \; .
\end{eqnarray}
Limits for zero arguments are useful. For $z \rightarrow 0$ we have
\begin{eqnarray}     \label{A500}
  j_l(z) \rightarrow  \frac{z^{l}}{(2l+1)!!}  \; \; , \;
  \eta_l(z) \rightarrow  \frac{(2l+1)!!}{(2l+1)} \frac{1}{z^{l+1}} \;.
\end{eqnarray}

\end{appendix}

\end{document}